\documentclass[mathleft
]{an}
\usepackage{graphicx}
\usepackage{times}
\begin{document}

% The following seven commands are intended for editorial usage and should be ignored by
% the author(s).
\Pagespan{1}{5}% Document's page range.
% If second parameter is left empty, the last page is computed automatically.
\Yearpublication{2007}%
\Yearsubmission{2007}%
\Month{0}%
\Volume{0}%
\Issue{0}%
% \DOI{This.is/not.aDOI}%

\title{Magnetic field evolution in neutron stars}

\author{Andreas Reisenegger\inst{1,2}}
%Example
%for footnote, note the usage of the \texttt{fnmsep}
%command as separator between institute number and footnote mark}
%\and  G.H. Ostwriter\inst{2,3}
%}
\titlerunning{Neutron star magnetic fields}
\authorrunning{Reisenegger}
   \institute{Max-Planck-Institut f\"ur Astrophysik, Karl-Schwarzschild-Str. 1,
             85741 Garching bei M\"unchen, Germany\\
         \and
             Departamento de Astronom{\'\i}a y Astrof{\'\i}sica,
             Pontificia Universidad Cat\'olica de Chile, Casilla 306, Santiago 22,
             Chile\thanks{Permanent address}\\
              \email{areisene@astro.puc.cl}\\
             }

\received{} \accepted{} \publonline{}


\abstract{%
Neutron stars contain persistent, ordered magnetic fields that are
the strongest known in the Universe.
%, going up to at least $\sim 10^{15}~\mathrm{G}$.
However, their magnetic fluxes are similar to those in magnetic A
and B stars and white dwarfs, suggesting that flux conservation
during gravitational collapse may play an important role in
establishing the field, although it might also be modified
substantially by early convection, differential rotation, and
magnetic instabilities. The equilibrium field configuration,
established within hours (at most) of the formation of the star,
is likely to be roughly axisymmetric, involving both poloidal and
toroidal components. The stable stratification of the neutron star
matter (due to its radial composition gradient) probably plays a
crucial role in holding this magnetic structure inside the star.
The field can evolve on long time scales by processes that
overcome the stable stratification, such as weak interactions
changing the relative abundances and ambipolar diffusion of
charged particles with respect to neutrons. These processes become
more effective for stronger magnetic fields, thus naturally
explaining the magnetic energy dissipation expected in magnetars,
at the same time as the longer-lived, weaker fields in classical
and millisecond pulsars.}

\maketitle

\section{Introduction}
The purpose of this talk is to present and discuss some of the
physical processes that are likely to be relevant in determining
the structure and evolution of magnetic fields in neutron stars,
almost regardless of their (so far largely unknown) internal
composition and state of matter. For this reason, I do not discuss
the fascinating, exotic issues of quark matter, superfluidity,
superconductivity, and the like, but emphasize the much more
pedestrian concepts of stable stratification and non-ideal
magnetohydrodynamics (MHD) processes such as ambipolar diffusion.
A review with a very different focus has recently been given by
Geppert (2006).

\section{Charged particles in neutron stars}
The name ``neutron stars'' incorrectly suggests stars composed
exclusively of neutrons. However, additional particles inside
these stars play a crucial role. A neutron ($n$) in vacuum is
known to decay by the weak interaction process $n\to
p+e+\bar\nu_e$ (beta decay) into a proton ($p$), an electron
($e$), and an electron antineutrino ($\bar\nu_e$), with a
half-life close to 15 minutes. This is impeded in very dense
matter by the Pauli exclusion principle: If all the low-energy
proton and electron states are already occupied, the only neutrons
that can decay will be those with a sufficiently high energy. On
the other hand, if there are many protons and electrons present,
some of these will be energetic enough to combine into neutrons by
inverse beta decay, $p+e\to n+\nu_e$, where $\nu_e$ stands for an
electron neutrino. In a neutron star, the neutrons and protons
will be confined by gravity, the electrons by the electrostatic
potential of the protons (see, e.~g., Reisenegger et al. 2006),
while neutrinos and antineutrinos are unbound and escape,
contributing to the cooling of the star (e.~g., Yakovlev et al.
2001).
%; Yakovlev \& Pethick 2004).
Direct and inverse beta decays
will be in balance if the chemical potentials\footnote{At zero
temperature, these chemical potentials reduce to the respective
Fermi energies.} of neutrons ($\mu_n$), protons ($\mu_p$), and
electrons ($\mu_e$) satisfy the relation $\mu_n=\mu_p+\mu_e$,
which forces the coexistence of a small fraction (few percent, but
density-dependent) of charged particles with a much larger number
of neutrons (e.~g., Shapiro \& Teukolsky 1983). Additional
particles (both charged and uncharged) can appear by other weak
decay processes at densities higher than typical nuclear
densities.

In addition to stabilizing the neutrons, the charged particles
play two important roles regarding the magnetic fields and their
evolution:
\begin{itemize}
\item Being charged, these particles can generate electrical
currents, which support potentially very strong magnetic fields.
\item Since the proton fraction $Y\equiv n_p/n$ depends on density
($n_i$ stands for the number density of particle species
$i=n,p,e$, and $n\equiv n_n+n_p$ is the total baryon density),
neutron star matter is inhomogeneous, stabilizing it with respect
to convective overturn (Pethick 1992; Reisenegger \& Goldreich
1992; Reisenegger 2001a). As discussed below, this is likely to
have an important stabilizing effect on magnetic field
configurations.
\end{itemize}

\section{Magnetic fields in neutron stars}
Many neutron stars are detected as pulsars, whose regular
pulsations in the radio, X-ray, and/or optical bands are produced
by a strong magnetic field being turned around at the stellar
rotation period $P$. These periods slowly increase in time, i.~e.,
the neutron stars lose rotational energy, probably through
magnetic coupling with their surroundings. Modelling this coupling
as electromagnetic radiation from a dipole rotating in vacuum,
oriented orthogonally to the rotation axis, one can infer the
surface magnetic field strength $B\propto\sqrt{P\dot P}$, where
$\dot P$ is the time-derivative of the rotation period (e.~g.,
Shapiro \& Teukolsky 1983). Inferred fields range from
$10^8~\mathrm{G}$ in millisecond pulsars up to
$10^{15}~\mathrm{G}$ in soft gamma-ray repeaters (SGRs; Duncan \&
Thompson 1992; Kouveliotou et al. 1998; Woods et al. 1999), the
latter being the strongest magnetic fields known in the Universe.

In radio pulsars, the rotational energy loss can account for the
whole observed energy output (relativistic particles and
electromagnetic radiation) from these objects. For the strongly
magnetized, but slowly rotating SGRs and anomalous X-ray pulsars
(AXPs), however, the observed X-ray luminosity is much larger than
the rotational energy loss rate, so an additional source of energy
is required, the most likely being the decay of their magnetic
field (Thompson \& Duncan 1996). This would make these objects be
the only known magnetically powered stars, or ``magnetars''. An
interest way of probing the strong magnetic fields inside these
objects appear to be the quasi-periodic oscillations recently
detected following two large flares of SGRs and interpreted as
crustal shearing modes coupled to Alfv\'en waves travelling
through the stellar core (Levin 2007).

In very old neutron stars, such as millisecond pulsars and
low-mass X-ray binaries, the magnetic field is $<10^9\mathrm{G}$,
weaker than in young neutron stars, such as radio pulsars and
high-mass X-ray binaries ($\sim 10^{11-14}\mathrm{G}$), suggesting
that the magnetic field strength decays with time. However, this
decay may be induced by accretion of matter from the binary
companion (e.~g., Payne \& Melatos 2007 and references therein)
rather than being a process intrinsic to the neutron star, so we
do not consider it further.
% in this presentation.

Magnetic field decay within the population of single %, classical
radio pulsars has also been suggested by some authors (Ostriker \&
Gunn 1969; Narayan \& Ostriker 1990) but disputed by others
(Bhattacharya et al. 1992; Faucher-Guigu\`ere \& Kaspi 2006), and
does not seem to be well established.

Aside from its intrinsic importance, the magnetic field may also
play an indirect role in deforming the neutron star and in this
way producing precession (Wasserman 2003), as appears to be
observed in some pulsars, and gravitational waves, which might
quickly reduce the rotation rate of newborn neutron stars (Cutler
2002).

\section{Origin of the magnetic field}

A natural hypothesis to explain the origin of the strong fields
observed in neutron stars is the compression of the magnetic flux
already present in the progenitor stars. It led Woltjer (1964) to
predict field strengths of $10^{15}~\mathrm{G}$, before any
neutron stars had been identified observationally. Later, it was
pointed out by many authors (e.~g., Ruderman 1972; Reisenegger
2001b; Ferrario \& Wickramasinghe 2005a,b, 2006) that the
distribution of magnetic fluxes is very similar in magnetic A and
B stars, white dwarfs, and neutron stars, in this way providing
support for the hypothesis of the magnetic fluxes being generated
on or even before the main-sequence stage and then inherited by
the compact remnants.

On the other hand, Thompson \& Duncan (1993) pointed out that
newborn neutron stars are likely to combine vigorous convection
and differential rotation, making it likely that a dynamo process
might operate in them. They predicted fields up to
$10^{15-16}~\mathrm{G}$ in neutron stars with few-millisecond
initial periods, and suggested that such fields could explain much
of the phenomenology associated with SGRs and AXPs (Duncan \&
Thompson 1992; Thompson \& Duncan 1995, 1996), some of which were
later confirmed to spin down at a rate consistent with a strong
dipole field ($10^{14-15}\mathrm{G}$; Kouveliotou et al. 1998;
Woods et al. 1999).

Of course, the two processes are not mutually exclusive. A strong
field might be present in the collapsing star, but later be
deformed and perhaps amplified by some combination of convection,
differential rotation, and magnetic instabilities (Tayler 1973;
Spruit 2002). The relative importance of these ingredients depends
on the initial field strength and rotation rate of the star. For
both mechanisms, the field and its supporting currents are not
likely to be confined to the solid crust of the star, but
distributed in most of the stellar interior, which is mostly a
fluid mixture of neutrons, protons, electrons, and other, more
exotic particles. For this reason, the present discussion focuses
on such an environment.

\section{Persistent, ordered field structures}

The magnetic fields of neutron stars, like those of upper main
sequence stars and white dwarfs, appear to be ordered (with a
roughly dipolar external configuration) and persistent (for much
longer than a solar cycle, perhaps for the en\-tire existence of
these stars). As mentioned above, their magnetic flux
distributions are similar, which also implies that their ratios of
magnetic to gravitational energy (or magnetic stress to fluid
pressure) cover a similar range, up to $\sim 10^{-6}$. Thus, even
the strong fields of magnetars are fairly unimportant in terms of
affecting the structure of the star. (The presence of
superconducting protons might somewhat increase this ratio, but
only if their critical field is substantially larger than the
actual field, which does not seem likely in the case of
magnetars.)

Another shared property of these stars is that a large part (if
not all) of their interior is stably stratified; in the case of
upper main-sequence envelopes and white dwarfs, because of a
radially increasing entropy; in the case of neutron stars, because
of a radial dependence in the fraction of protons, electrons, and
possibly other particles.

This makes it natural to search for magnetic field configurations
that might be stable in stably stratified stars. Analytic attempts
in the past have failed, only yielding the general result that
both purely toroidal fields and purely poloidal fields are
unstable (Tayler 1973; Flowers \& Ruderman 1977), and the
speculation that linked toroidal and poloidal fields might
stabilize each other, yielding a stable equilibrium (Prendergast
1956; Wright 1973).

Recently, Braithwaite \& Spruit (2004, 2006; see also Braithwaite
\& Nordlund 2006) carried out MHD simulations of stars with a
stabilizing entropy gradient, which were initially given complex,
``random'' magnetic fields. These were observed to evolve on an
Alfv\'en-like timescale into a linked poloidal-toroidal
configuration that persisted for a dissipative timescale. Thus,
this configuration might be a good approximation to the actual
field structures in upper-main sequence, white dwarf, and neutron
stars. These would be stable in ideal MHD, but evolve over long
times due to various dissipative processes present in each kind of
star.

In what follows, we will argue that the stable stratification of
the stellar matter is crucial in allowing stable magnetic
equilibria. In neutron stars, the compositional stratification can
be overcome by processes that either move different particle
species with respect to each other (ambipolar diffusion) or turn
them into each other (beta decays). Both processes are enhanced by
strong magnetic fields, yielding a natural explanation for the
field decay in magnetars and its absence in other isolated neutron
stars.

\subsection{Single flux tube as a toy model}

In order to understand the importance of stable stratification for
the equilibria and evolution of the magnetic field, consider a
horizontal magnetic flux tube with field strength $B$, fluid
pressure $P_\mathrm{in}$, and fluid density $\rho_\mathrm{in}$
embedded in a plane-parallel, unmagnetized fluid with pressure
$P_\mathrm{out}$, and density $\rho_\mathrm{out}$, stratified by a
gravitational field. In order to be in mechanical equilibrium with
its surroundings, it must satisfy
$P_\mathrm{in}+B^2/{8\pi}=P_\mathrm{out}$ and
$\rho_\mathrm{in}=\rho_\mathrm{out}$. In a barotropic fluid, with
a unique pressure-density relation $\rho(P)$, these two relations
cannot be satisfied simultaneously. If the first is satisfied,
then $P_\mathrm{in}<P_\mathrm{out}$ and therefore
$\rho_\mathrm{in}<\rho_\mathrm{out}$, so the flux tube will
buoyantly rise (Parker 1974). However, in neutron star matter,
since the composition (here represented by the proton fraction
$Y$) is inhomogeneous, the equation of state has the form
$\rho(P,Y)$. Thus, the flux tube can move to a position where a
composition difference $\delta Y\equiv
Y_\mathrm{in}-Y_\mathrm{out}\sim 3B^2/(8\pi n\mu_e)$ compensates
for the pressure difference, allowing the densities to be equal.
So, the flux tube remains in equilibrium over time scales on which
the matter can be modelled as a single fluid.

Over longer time scales, two effects can erode the equilibrium
described above. On the one hand, the matter inside the flux tube
will be out of beta equilibrium, and this equilibrium can be
restored by weak interaction processes such as the neutron beta
decay or its inverse. On the other hand, particles can move in and
out of the flux tube, in this way also equalizing the internal and
external composition. On the characteristic time scale for these
processes (which is very long, due to the strong degeneracy and
particle collisions), the flux tube will move a distance over
which the external composition changes by $\delta Y$ as defined
above. This implies that the time scale for the flux tube to move
a scale height is $\sim Y/\delta Y\approx 8\pi n_e\mu_e/(3B^2)$
times longer than the shorter of the beta equilibration and
ambipolar diffusion times in the flux tube.

In a more realistic geometry, one might consider a toroidal flux
tube in the equatorial plane of a spherical star. In this case, in
addition to buoyancy, which tends to make it move radially
outwards, the tube will be subject to a tension force, which
exerts the opposite effect, namely to contract inward, towards the
axis. At large radii, the former will be stronger, while the
latter dominates at small radii. As for the plane-parallel case,
this motion will actually happen in a barotropic star, but it will
soon stop in one with a strong, stable stratification. In the
latter, it can move only very little in the radial direction, so
further motion is constrained to happen essentially on spherical
surfaces. Thus, the flux tube will be unstable to contract on a
spherical surface.

\subsection{Implications for stable equilibria}

As illustrated by the example of the single flux tube, the
magnetic flux in the interior of a realistic, stably stratified
neutron star can only move radially on long time scales controlled
by weak interactions and ambipolar diffusion (as well as the Hall
drift). In a combined toroidal-poloidal field configuration such
as that found by Braithwaite \& Spruit (2004, 2006; also
Braithwaite \& Nordlund 2006), the presence of the poloidal field
in the region around the magnetic axis impedes the toroidal field
from contracting onto the latter. At the same time, the poloidal
field cannot deform significantly, because it is ``tied together''
by the toroidal field. Thus, it appears natural that such
equilibria do exist in stably stratified stars, but do not exist
in barotropic ones.

Additional insight can be reached through another, more formal
argument: The combined pressure and gravitational force on a fluid
element can be written as
\begin{equation}\label{force}
\vec f=-\nabla\delta P+\delta\rho~\vec g,
\end{equation}
where $\vec g$ is the gravitational acceleration, and $\delta P$
and $\delta\rho$ are small, Eulerian perturbations of pressure and
density. If the whole configuration is axisymmetric, there will be
no fluid force components in the azimuthal ($\phi$) direction,
thus the magnetic force in this direction must also vanish, $\vec
j\times\vec B\cdot\hat\phi=0$. The ``allowed'' magnetic field
configurations will therefore produce two independent force
density components, in the poloidal plane, which must be cancelled
by two components of the fluid forces of eq.~(\ref{force}). This
is generally possible in a stably stratified fluid, where $\delta
P$ and $\delta\rho$ are independent functions of position. In a
barotropic fluid, however, these two functions are proportional to
each other, and the allowed equilibria will be much more
restricted, if they exist at all. Thus, in a stably stratified
fluid, there will be many equilibria, some of them likely stable,
whereas in a barotropic fluid there will be few, if any.

\subsection{Implications for field evolution}

On the same time scales on which a single flux tube can move
radially (controlled by weak interactions and ambipolar
diffusion), the belt of twisted toroidal field lines in the
Braithwaite-Spruit configurations can also move outwards,
decreasing the magnetic energy stored in the star. However, it was
shown above that the time scale for this to happen is very long.

In fact, for the case in which weak interactions provide the
dominant dissipation, the time scale to reach beta equilibrium is
at best a few times shorter than the cooling time of the
star\footnote{Both beta equilibration and cooling depend on weak
interactions and have the same temperature-dependence, until the
star is cool enough for photon cooling to become dominant.}
(Reisenegger 1995), unless the chemical imbalance
$\eta\equiv\mu_n-\mu_p-\mu_e\approx B^2/(8\pi n_e)$ is very large,
$|\eta|\gg kT$. Since a substantial motion of the magnetic
structure takes $\sim Y/\delta Y\approx 8\pi n_e\mu_e/(3B^2)\sim
(4\times 10^{17}\mathrm{G}/B)^2$ times the equilibration time, the
field would never decay, were it not because the field decay
itself could keep the star hot for a longer time (Thompson \&
Duncan 1996; Reisenegger et al. 2005). A strong field will remain
essentially constant until the temperature has decreased to
$kT\approx|\eta|/5.5$, at which point the weak interaction
processes start having a net heating effect (Fern\'andez \&
Reisenegger 2005), converting the energy stored in the chemical
imbalance into neutrinos and heat. Thus, the stellar interior
remains essentially at this level until the magnetic field has
decayed substantially (Thompson \& Duncan 1996).

The ambipolar-diffusion-dominated limit is relevant at lower field
strengths (the transition occurs near $10^{16}~\mathrm{G}$), and
here again the heat input from the magnetic field can eventually
offset the cooling and keep the star warm until the field has
decayed.

The time scales for these processes are difficult to estimate
precisely (and may depend strongly on Cooper pairing effects,
neutrino emission mechanisms, and the like), but estimates based
on Goldreich \& Reisenegger (1992) appear to be roughly consistent
with the time scales $\sim 10^{4-5}~\mathrm{yr}$ required to
account for the magnetar phenomenon.

\section{Crossing the solid crust}

Of course, neutron stars have a solid crust that must be crossed
by the magnetic flux on its way out of the star. Its precise
properties are still not clear, as it is a very unusual solid,
subject to a strong pressure, strongly stratified, and with a
relatively weak shear modulus. Thus, it is not clear whether
strong magnetic stresses would crack the crust in discrete events,
suddenly releasing a substantial amount of energy and possibly
accounting for the soft gamma-ray bursts (Thompson \& Duncan
1996), or rather flow plastically, with a continuous release of
energy (Jones 2003). If the magnetic forces are not strong enough
to break and move the crustal solid, the flux will have to be
transported by a combination of Ohmic diffusion and Hall drift
(e.~g., Jones 1988; Goldreich \& Reisenegger 1992; Pons \& Geppert
2007; Reisenegger et al. 2007).

\section{Conclusions}

The extremely strong magnetic fields found in neutron stars are
nevertheless weak enough to be balanced by small perturbations in
the density, pressure, and chemical composition of the stably
stratified, degenerate, multi-species fluid found in their
interior. The field likely originates from the fairly strong
magnetic flux inherited from the progenitor star, possibly
modified by a combination of convection, differential rotation,
and magnetic instabilities acting during the short, protoneutron
star stage following collapse. The equilibrium field set up during
this stage likely involves linked toroidal and poloidal field
components. It can decay through dissipative processes such as
weak interactions and ambipolar diffusion, which change the
chemical composition of the matter, allowing it to move. These
processes become particularly effective at high field strengths,
possibly accounting for the energy release inside magnetars, which
also keeps the stellar interior hot enough for the magnetic field
to rearrange substantially. This forces changes as well in the
crust of the star, which can be broken by strong fields, whereas
weaker ones can evolve by a combination of Hall drift and
resistive dissipation. The internal dissipation processes are
ineffective for pulsar-strength fields, which can easily survive
for a pulsar lifetime. None of these processes depends essentially
on the exotic properties of a neutron star, such as Cooper pairing
or quark matter, which can however modify the time scales for
these processes to occur.

\acknowledgements
      The author thanks P. Goldreich, Y. Levin,
      F. Meyer, H. Spruit, C. Thompson, A. Watts, and Y. Wu
      for many stimulating and informative conversations.
      This work was supported by FONDECYT (Chile)
      Regular Research Grant 1060644.
\newpage%%%%%%%%%%%%%%%%%%%%%%%%%%%%%%%%%%%%%%%%%%%%%%%%%%%%%%

\end{document}